\documentstyle[prl,aps]{revtex}

\begin{document}

\draft

\title{Response to Eisenstein and Bunn's Null Hypothesis 
Comment on Cosmological Birefringence}

\author{Borge Nodland}

\address{Department of Physics and Astronomy, and Rochester Theory
Center for Optical Science and Engineering, University of Rochester,
Rochester, NY 14627}

\author{John P. Ralston}

\address{Department of Physics and Astronomy, and Kansas Institute for
Theoretical and Computational Science, University of Kansas, Lawrence,
KS 66044}

\date{Submitted to Physical Review Letters (1997)}

\maketitle

\begin{abstract}
Eisenstein and Bunn have widely circulated a Comment (astro-ph/9704247)
suggesting a non-standard null hypothesis in a test for cosmological
birefringence. The Comment misrepresents the procedure used by Nodland
and Ralston (B. Nodland and J. P. Ralston, {\it Phys. Rev. Lett.} {\bf
78,} 3043 (1997); astro-ph/9704196) and lacks statistical basis; no
calculations were reported, but sweeping conclusions were drawn from
eyeballing a single scatter plot. The results of the suggested
procedure range from an underestimate of the statistical significance
of well--correlated data, to a failure of detecting a perfect
correlation in the limit of strong cuts. We verified these faults by
performing the actual calculation which EB suggested, but neglected to
carry out themselves.  Furthermore, the calculation showed that the
original correlation remains statistically significant, with the
lego--plot of $1/P$ versus $\vec{s}$ showing a persistent bump, and the
anisotropy direction $\vec{s}$ remaining in the same direction as
previously reported.
\end{abstract}              


Eisenstein and Bunn's Comment \cite{one} is weak in making sweeping
judgments without much foundation, and with absolutely no calculation.
It is unfortunate that the Comment misrepresents what we reported in
\cite{two}, and we find it flat wrong in several assertions. We
welcome the opportunity to clear up the matter.

Our angle $\beta$ is not so simply defined as ``the angle between the
polarization direction and the galaxy axis,'' but is a four--part
formula involving the fundamental variables $\chi$ and $\psi$, the
galaxy position, and the trial axis $\vec{s}$. In contrast to variables
used in other studies, the angle $\beta$ retains information about the
sense of rotation being tested. We did not pick ``the angle $\beta$
from a uniform distribution of allowed angles.'' Instead, and as
reported in our paper, we picked the $\chi$'s and $\psi$'s from a
uniform distribution. The distributions of $\chi$'s and $\psi$'s found
in the data {\it are} quite uniform, so our Monte Carlo reproduces what
the data shows. We then fed the random $\chi$'s and $\psi$'s into the
same formula that evaluated $\beta$ for the data. The result is that
the $\beta$ distribution in each quadrant depends on everything,
including the anisotropy of the galaxy distribution and the trial
$\vec{s}$. The distribution of random $\beta$'s is whatever is chosen
by the procedure 1 or procedure 2 Monte Carlos, and is not what
Eisenstein and Bunn (EB) describe.

In the fifth paragraph in \cite{one}, EB make a claim that is
unsupportable: ``if the underlying galaxy distribution truly had a
uniform intrinsic distribution in $\beta$, it would be impossible to
measure the proposed birefringence at all.'' This is incorrect, as
every measurement that obtains a set of data points lying along the
line $y = m x + b$ can establish a correlation between $y$ and $x$,
whatever the distribution of data sampled on the $y$--axis. For
example, if the $\beta$ versus $r \cos \gamma$ plane had points equally
spaced and exactly along the diagonal line, the distribution over
$\beta$ would be uniform, but with $\beta$ and $r \cos \gamma$
obviously correlated.

We think that what EB is trying to say is in the next paragraph, where
``estimating by eye'' as they put it, the data in Fig. 1(d) in \cite{two}
are observed to be ``more tightly correlated than data uniformly
distributed between $0$ and $\pm\pi$.'' This is put forth as something
bad, or suspicious, without recognizing that there is a very good
reason for the $\beta$ values to be distributed this way. The figure
shows data surviving the cut $z \geq 0.3$, which leaves a hole on both
sides of the origin on the $r \cos \gamma$ axis. This was described in
our paper. The data is correlated like $\beta = r \cos \gamma$, and
naturally the $\beta$ values are going to have a hole at the origin of
the $\beta$ axis!  What about the outer ends?  We run out of data at
large $z$, so $r \cos \gamma$ has a maximum absolute value. Again,
because the data is correlated, one indeed finds a shortage of
$\beta$'s of large magnitude. Isn't it true that whenever there is data
on the ``$y$--axis'' well correlated with data on the ``$x$--axis,''
one will find that the ``$y$--axis'' distribution reflects the cuts and
limits found on the ``$x$--axis''?

EB thus make a wrong deduction about the underlying distribution of
$\beta$ values. To account for their own conclusion, EB suggest for
their own faux null hypothesis to ``draw the angles ($\beta$) from the
observed distribution'' and make an estimate of the significance
relative to that. This has the sound of something fancy, but what is
the basis for it, and what does it really mean? Suppose one has
laboratory data that exhibits a linear relation of the form $y = m x$,
plotted on a diagonal line. With $x$--cuts like ours, the data is
restricted to boxes in two quadrants. If one uses that data and
shuffles it, one makes a set of faux random data coming uniquely from
the small boxes. This ``random data'' is far from random; it is highly
pre--correlated. Comparing the correlations from random shufflings with
the better correlated real data, one will find a correlation, but the
baseline for what is declared ``relatively likely'' then gets moved
up.  The generic results are that the data's correlation can be
artificially made to look more probable. For example: the significance
of even a (perfectly correlated) $\delta(y - x)$ distribution, when
compared with such well--correlated faux randoms, would be
meaningless in the limit of strong cuts, with a probability of unity
for the faux randoms to be as correlated as the true $x$ and $y$.

It seems to us that EB did not understand our procedure, and created a
hasty argument based on a false premise from eyeballing a single
scatter plot. Their opinion that we overestimated the statistical
significance of our results is incorrect. The description could also
mislead a reader who might not know that we did not quote any
``absolute'' probabilities based on statistics books, but restricted
ourselves to reporting just what we found: the P--values for clearly
defined and objective Monte Carlo procedures that were done two
separate ways. Anyone who wants to do the work can go ahead and make a
serious calculation based on some other criterion any time.

To back up our claims, we checked our belief that the EB procedure will
artificially raise the baseline of a good correlation. Repeating our
calculation with shuffling that kept the $\beta$ distribution
invariant, we found that the P--values were increased, with the
smallest P--values (those forming the peak in the $1/P$ plot in Fig. 1(b)
in \cite{two}) increased by a factor of about ten above our previously
reported values. Even with this change (which we don't accept as
meaningful), the correlation still looks significant. The lego--plot
showing a bump in $1/P$ versus $\vec{s}$ [our Fig. 1(b)] still showed a
bump with the ``EB procedure,'' and the bump occured for the same direction
of $\vec{s}$ as before. We went ahead and did the calculation, because
we think it would be bad science to make general remarks in the absence
of attention to the facts.

In a final complaint, EB object to the use of the scale $\Lambda_s$,
which they claimed to be inappropriate because it may be non--zero for
random data. The facts are, we did not use $\Lambda_s$ as a statistic,
but used the correlation coefficients $R$. EB might have considered the
fact that every curve--fitting or statistical Monte Carlo procedure can
generate a non--zero parameter for random data. The non--zero
parameters are ignored when the correlation is not significant. In our
analysis, the Monte Carlo test rejected the null hypothesis as a
consequence of P--values being of order $10^{-3}$. When the test
rejects the null hypothesis, calculation of $\Lambda_s$ is a meaningful
thing to do, just as we reported. 

Sweeping judgments can be dangerously wrong, and, when widely
distributed without responsibility, can harm good work honestly
reported.

\end{document}